# Magnon Transport in the Presence of Antisymmetric Exchange in a Weak Antiferromagnet

A. Ross[a,b], R. Lebrun[c], O. Gomonay[a], J. Sinova[a], A. Kay[d], D. A. Grave[d,1], A. Rothschild[d], M. Kläui[a,b,e*]

  a. Institut für Physik, Johannes Gutenberg-Universität Mainz, D-55099, Mainz, Germany.
  b. Graduate School of Excellence Materials Science in Mainz (MAINZ), Staudingerweg 9, D-55128, Mainz, Germany.
  c. Unité Mixte de Physique, CNRS, Thales, Université Paris-Saclay, Palaiseau, 91767, France.
  d. Department of Materials Science and Engineering, Technion-Israel Institute of Technology, Haifa 32000, Israel.
  e. Center for Quantum Spintronics, Department of Physics, Norwegian University of Science and Technology, Trondheim, Norway.

  *klaeui@uni-mainz.de

**Abstract**

The Dzyaloshinskii-Moriya interaction (DMI) is at the heart of many modern developments in the research field of spintronics. DMI is known to generate noncollinear magnetic textures, and can take two forms in antiferromagnets: homogeneous or inter-sublattice, leading to small, canted moments and inhomogeneous or intra-sublattice, leading to formation of chiral structures. In this work, we first determine the strength of the effective field created by the DMI, using SQUID based magnetometry and transport measurements, in thin films of the antiferromagnetic iron oxide hematite, $\alpha\text{-}Fe_2O_3$. We demonstrate that DMI additionally introduces reconfigurability in the long distance magnon transport in these films under different orientations of a magnetic field. This arises as a hysteresis centred around the easy-axis direction for an external field rotated in opposing directions whose width decreases with increasing magnetic field as the Zeeman energy competes with the effective field created by the DMI.

**Main Text**

### 1. Introduction

Magnonic excitations in magnetic systems have recently received significant interest due to various possible applications in the field of magnonics[1]. In particular with the advent of low dissipation

Present Address:
1. Department of Materials Engineering and Ilse Katz Institute for Nanoscale Science and Technology, Ben-Gurion University of the Negev, Beer-Sheva 8410501, Israel.

magnetic insulators, spin transport by magnonic spin currents have been investigated with exciting findings such as long distance spin transport in the ferrimagnetic garnet YIG ($Y_3Fe_5O_{12}$)[2,3]. In particular thermal magnonic spin currents due to the spin Seebeck effect[4] have been observed experimentally[5] and described theoretically with important contributions by S. M. Rezende et al.,[6]. More recently the theory was extended to antiferromagnetic insulators[7,8], as these materials exhibit a distinctly different magnon spectrum and are not susceptible to stray fields[9,10]. An excellent introduction to magnons in antiferromagnets can be found in Ref. [11].

A particularly exciting antiferromagnetic insulator is the antiferromagnetic phase of iron oxide α-$Fe_2O_3$, known as the mineral hematite, which is employed for a variety of diverse applications. Recently, this list has been extended to include being a key front runner for antiferromagnetic magnonic devices, demonstrating micrometer spin diffusion lengths[12–17]. When cooling to below the Néel temperature, the staggered magnetization *n* lies in the (0001)-plane where an emergent canted moment *m* appears perpendicular to *n* due to a bulk effective field $H_D$ from the Dzyaloshinskii-Moriya interaction (DMI) directed along [0001][18]. This material undergoes a rotation of *n* to orient along [0001] (see Fig. 1a) and *m* disappears below the Morin temperature $T_M$[19–21]. Below this transition temperature, $H_D$ modifies the effective anisotropy field, affecting the external magnetic field **H** required for reorienting *n*[20–22]. Above $T_M$, the [0001] axis is the hard-magnetic axis, whilst below $T_M$, the [0001] direction is the magnetic easy-axis. In the easy-axis phase, the excited magnons are circularly polarized and a net transport of angular momentum is possible by altering the relative populations of the lowest magnon modes[12,13]. The transport in the high temperature easy-plane phase above $T_M$, where the magnon modes are linearly polarized, is mediated instead through a superposition of the available modes that dephase over some length scale[14–16]. Above $T_M$, the mechanisms behind the magnon transport of angular momentum has been related to the relative strength of $H_D$[16,17] but good agreement has also been found without considering it[14,15,17]. Meanwhile below $T_M$, explanations of the magnon transport have not considered the effect of the DMI outside of aiding the rotation of $\boldsymbol{n}$[12,13] raising the question of whether the DMI is relevant or even crucial in the magnon transport mechanisms investigated so far.

In this work, we first extract the strength of $H_D$ in thin films of hematite through measurements of the canted moment and spin transitions, demonstrating that there still exists a sizeable $H_D$. We then find that the transport of magnons in the easy-axis phase of hematite is affected by this additional DMI competing with the Zeeman energy under an applied field, resulting in hysteretic magnon transport when this magnetic field is rotated.

We use 500 nm thin films of hematite grown on sapphire substrates by pulsed laser deposition[23,24], orientated as ($1\bar{1}02$), which is known as r-plane orientation. The unit cell of hematite is shown in Fig. 1a with the r-plane indicated, showing the 33° angle made between the surface plane and the [0001] axis. Making use of electron beam lithography and the deposition of 7 nm thick Pt, Hall bars and non-local wires were produced for the electrical measurements[12,13,20].

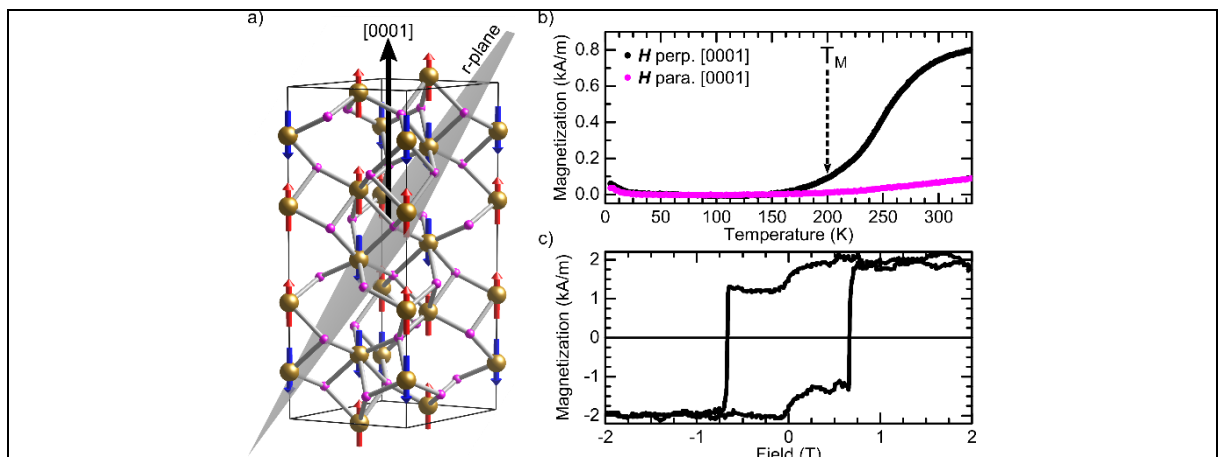

**Figure 1**: **Magnetization of thin film hematite**. a) Unit cell of hematite (α-$Fe_2O_3$) in the easy-axis anisotropy phase below $T_M$. The [0001] axis is indicated, and the r-plane is shown in grey. The gold atoms are Fe atoms and pink are O atoms. The red/blue arrows represent the magnetic moments[25]. b) Magnetization of 500 nm, r-plane hematite thin films as a function of temperature for a magnetic field within the (0001)-plane (black) or perpendicular to this plane (magenta). c) Magnetization of

500 nm r-plane hematite thin films versus field at 300 K measured by SQUID. A linear background has been subtracted to account for the diamagnetic substrate.

## 2. Results and Discussion

We start by investigating the effect of $H_D$ on the magnetic properties of our thin films making use of a superconducting quantum interference device (SQUID). The films were installed such that a magnetic field *H* is applied within the (0001) plane and we are sensitive to the net moment *m* originating due to DMI above $T_M$[26,27]. The relative magnitude of *m* stems from a competition between $H_D$, which favors a perpendicular alignment of the spins on opposing antiferromagnetic sublatttices, and the exchange field $H_E$, which favors an antiparallel alignment. By confirming and measuring the magnitude of *m* we can therefore determine the strength of the internal effective field present. After applying a large magnetic field of 3 T at high temperatures, the films were cooled in a probing field of 50 mT as the net magnetization was measured (Fig. 1b). The films were then warmed back to 330 K and no thermal hysteresis is observed. A temperature independent background (calculated as the average signal between 50 K and 100 K) is subtracted to account for the diamagnetic $Al_2O_3$ substrate and sample mounting contributions. The reduction of the magnetization that occurs due to the Morin transition is clearly visible (see Fig. 1b, black points), enabling us to establish the purely easy-axis phase exists below about 200 K. We label this temperature as the end of the Morin transition, $T_M$. This is confirmed by measuring with a magnetic field parallel to [0001], where no net moment appears (Fig. 1b, magenta points), and we observe no significant changes in the magnetization across the investigated temperature range. Fixing the temperature at 300 K, above $T_M$, we probe the magnetization as a function of *H* applied within the (0001)-plane. To account for the substrate and sample mounting contributions, we make an average linear fit to the signal above and below ±1 T and subtract this. We observe in Fig. 1c a clear hysteresis due to the canted moment with a coercive field of 690 mT ± 30 mT with a sharp switching due to the inversion of the canted moment *m*. We also note a soft loop at very low magnetic fields due to a thin ferrimagnetic seed layer that nucleates at the hematite/substrate interface[17,28]. The films are 500 nm thick, so this interfacial layer plays no role in the subsequent measurements including the transport data. We can then assert that the saturation magnetisation $m_s$ due to the canted moment is 2.1 ± 0.1 mT. From this, we can calculate the DMI effective field as $H_D = m_s/\chi$ = 1.7 ± 0.1 T, which is lower than the value extracted for bulk hematite[20]. This assumes that the antiferromagnetic susceptibility $\chi$ perpendicular to the [0001] direction is the same as in bulk hematite[27]. Such an assumption is valid given that the exchange field should be similar between bulk and 500 nm thick films as the Néel temperatures are similar.

Having established a value for the effective DMI field above $T_M$, we now turn our attention to extracting $H_D$ from surface-sensitive spin Hall magnetoresistance (SMR) measurements as previously reported for hematite and other AFMs[20,21,29–31]. The SMR is a powerful technique for detecting the spin transition of antiferromagnetic insulators, both field driven[20,29] and temperature driven[32]. In the case of hematite, by determining the critical magnetic fields required for spin-flopping of the Néel vector, not only can the internal effective anisotropies be determined, but also the strength of the effective DMI field. Below $T_M$, $H_D$ not only lowers the effective anisotropy field, thus reducing the spin-flop field $H_{sf}$, but also induces a spin-reorientation of *n* when a magnetic field is applied at any angle to the [0001] direction[20,22]. This leads to two key critical fields, $H_{sf} = (H_K H_E - H_D^2)^{1/2}$ where $H_K$ and $H_E$ are the anisotropy field along [0001] and the exchange field respectively, and $H_{sf,DMI} = H_{sf}^2/H_D$, the critical field when *H* and *n* are perpendicular[20]. For any other angle ξ, the critical magnetic field will fall between these two extremes. As can be seen, the critical magnetic field for the spin-reorientation of *n* depends intrinsically on the strength of $H_D$. We adopt the geometry shown in Fig. 2a where the charge current $j_c$ of a Pt Hall bar is perpendicular to the in-plane projection of the [0001] axis. The Hall bar is produced through lithographic methods and the sputter deposition of 7-nm Pt in an Ar atmosphere. We probe the longitudinal Pt resistance at 175 K, making use of two pads separated by 55 μm, by averaging between positive and negative probing currents to remove any offsets. We then calculate the longitudinal SMR response as $\Delta R_{xx}/R_0 = (R-R_0)/R_0$ where $R_0$ is the zero-field resistance. The charge current passed through the Pt leads to an interfacial spin accumulation µ, polarized along *x*. The value of the SMR is given by the projection between the Néel vector and µ, i.e. $\propto 1-(\boldsymbol{n} \cdot \boldsymbol{\mu})^2$. By measuring the SMR for a magnetic field parallel to the in-plane [0001] projection, i.e. *H*||*x*, we observe an increase and plateau at $\mu_0 H(\xi = 33°) \approx 8$ T ± 0.5 T indicating a rotation of *n*[20,21,29]. Applying *H* along *y* similarly leads to an increase in the resistivity but no clear plateau up to 11 T. However, making use of later measurements (Fig. 3d), we can expect that the a magnetic field will overcome all internal anisotropies of any origin at a magnetic field of $\mu_0 H_{sf,DMI} \approx 12.5$ T ± 1 T, above which, the orientation of *n* will be dictated by the magnetic field. An increase is seen for both field geometries because the projection $\boldsymbol{n} \cdot \boldsymbol{\mu}$ decreases as *n* smoothly rotates away from the easy-axis. The critical magnetic field for *n* to rotate when *H* is applied at a finite angle, $H_{cr}(\xi)$ is a function of $H_{sf}$, $H_{sf,DMI}$ and $\xi$[20]. This means that, by experimentally measuring $H_{cr}$ for $\xi = 33°$ and $H_{sf,DMI}$, we can estimate the value of the spin-flop field $H_{sf}$ to be about 4.5 T ± 1 T. Given that the strength of the DMI is determined by the ratio of the critical field extrema ($H_D = H_{sf}^2 / H_{sf,DMI}$), this leads to a value for $H_D$ of 1.6 T ± 0.4 T, in excellent agreement with the value extract from magnetometry measurements in Fig. 1.

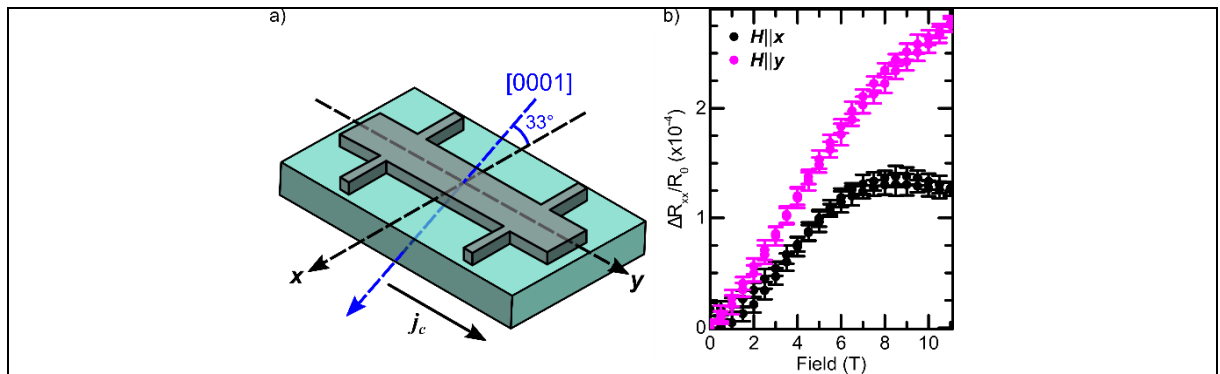

**Figure 2: Spin Hall magnetoresistance of r-plane thin film hematite.** a) Hall bar geometry on 500 nm r-plane thin film hematite. A charge current $j_c$ is passed perpendicular to the in-plane projection of the [0001] axis as a longitudinal resistivity is measured. b) Longitudinal spin Hall magnetoresistance of Pt/hematite bilayers at 175 K for a magnetic field applied parallel (black) or perpendicular (magenta) to the in-plane projection of the [0001] axis. $\Delta R_{xx}/R_0 = (R-R_0)/R_0$ where $R_0$ is the zero-field resistance.

Finally, having demonstrated that these films indeed have a significant bulk DMI, we will investigate the effect of this additional exchange path on the magnon transport, and demonstrate that the chiral nature of DMI can have a profound impact on the magnon transport efficiency. We adopt a lithographically produced non-local geometry of two 7 nm Pt wires deposited atop our hematite films[12–14,17], aligned perpendicular to the [0001] in-plane projection, see Fig. 3a. The wires are 80 μm long and 250 nm wide, with a center-to-center separation from 525 nm to over 1 μm. The application of a charge current through one wire leads to the excitation of spin-polarized magnons in the hematite films that propagate diffusively away[12]. The second Pt wire acts as a magnon absorber, converting the magnon spin current into a detectable voltage as previously demonstrated[12–14]. At an environment temperature of 175 K, we rotate a magnetic field within the sample plane through an angle α where α = 0° occurs for ***H*** parallel to the [0001] projection. Having rotated clockwise under +***H***, we invert the field and rotate anticlockwise. Given the 180° symmetry of ***n***, we would expect these two curves to overlap, even in the presence of DMI. However, surprisingly we observe a hysteresis emerging that is centred around 0°/180° between the two rotations (Fig. 3b). This hysteresis originates from the magnetic state of the hematite itself, affecting the magnon transport over large distances (Fig. 3c). The hysteresis appears due to a competition between the DMI energy $E_D$ and the Zeeman energy $E_Z$,

$$E_D \propto H_D \sin\theta (H_Y \cos\phi - H_X \sin\phi)$$

$$E_Z \propto \frac{1}{2}(H_X^2 \cos^2\phi + H_Y^2 \sin^2\phi)\sin^2\theta$$

which scale linearly and quadratically, respectively. In the above equations, $H_X$ and $H_Y$ are the components of the applied magnetic field appearing perpendicular to the [0001] direction and (θ,ϕ) are the polar coordinates of ***n***. We can see that, as soon as a finite angle appears between ***H*** and the [0001] direction, the DMI will begin to compete with the Zeeman response. Furthermore, the DMI energy depends on the direction of ***n*** and changes sign when the Néel vector is reversed by 180°. In other words, the DMI energy discriminates between states with opposing orientations of ***n***. By applying a magnetic field we set the relative orientation of the DMI term based on the history of the system. As we rotate the magnetic field within the sample plane through the in-plane orientation of [0001], the system resists the effect of the magnetic field strength, giving rise to the hysteresis. The size of the hysteresis (the deviation between the peak value in the vicinity of 0°/180° for each field polarity) is dependent on the applied magnetic field, and we observe that it decreases with increasing field following a 1/H dependence (Fig. 3d), supporting our conclusions that the magnon transport is affected by the additional DMI present in the system. The hysteresis is anticipated to experimentally disappear when $\mu_0 H$ = 14.9 T is applied along ***x***, corresponding to 12.5 T applied within the (0001) plane, the value we used earlier to extract $H_D$ from SMR measurements. It is worth noting also the different length-scales probed by local SMR measurements and non-local magnon transport. The SMR is sensitive predominantly to the magnetic state of the interface which represents the bulk magnetic state for antiferromagnets, with a penetration depth on the order of the exchange length, i.e. a few nanometers. Meanwhile, the magnon transport represents the magnetic state across a region on the order of the diffusion length, which for similar films is several hundred nanometers[13].

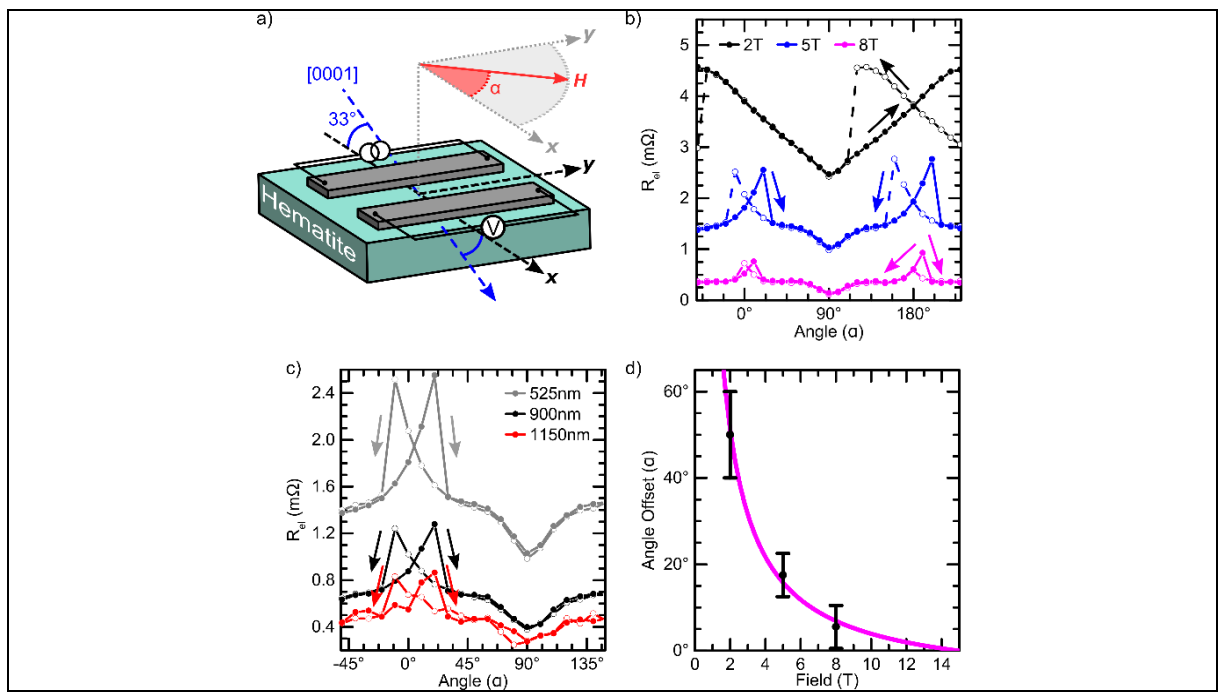

**Figure 3: Non-local magnon transport in r-plane thin film hematite at 175 K.** a) Non-local geometry of Pt wires atop 500 nm r-plane thin film hematite. The wires are aligned perpendicular to the in-plane projection of the [0001] axis. A charge current is passed through one wire and a voltage is measured in the second, allowing for a non-local resistance to be calculated. b) Non-local resistance for a magnetic field rotated in the sample plane for several magnetic fields where α = 0° occurs for *H*||*x*. The arrows indicate the direction of rotation for a positive field (filled symbols) or negative fields (empty symbols). c) Non-local resistance for a field of 5 T rotated in the sample plane for several wire separations where α = 0° occurs for *H*||*x*. The arrows indicate the direction of rotation for a positive field (filled symbols) or negative fields (open symbols). d) Angle between the peak of the hysteresis and 0°/180° versus field alongside a 1/*H* fit that goes to 0 at 14.9 T.

Making use of the methods described in Ref. [20], we can calculate the equilibrium orientation of the Néel vector under an applied magnetic field in the absence (i.e. $H_D$ = 0 T) or presence of a significant DMI field (Fig. 4). In the case of $H_D$ = 0 T, we find that the reorientation of the Néel vector under an applied magnetic field is identical under a reversal of the initial orientation of *n*, i.e. *n*$_{in}$ as shown in Fig.4a and Fig. 4b. The reversal of *n*$_{in}$ is identical to a reversal of *H*, and thus we would expect a completely symmetric signal under a reversal of field. However, in the case of a significant DMI, as the strength of the magnetic field is increased, we observe drastically different trajectories of the Néel vector reorientation (Fig.4c and 4d). In the case of *n*$_{in}$ lying along -*z*, we even observe a discontinuity in the position of *n* with increasing field. This demonstrates that the presence of an effective DMI field can have a profound impact on the equilibrium position of the Néel vector that is dependent not only on the strength of $H_D$ and *H*, but also on the initial conditions of *n* with respect to the field polarity. When we experimentally rotate the sample under an applied field, we then illicit differing trajectories for *n*, giving rise to a hysteresis in the equilibrium position, and thus a hysteresis in the magnon transport. It is also worth bearing in mind that thin film antiferromagnets can have a multidomain state that has been shown to have an impact on the non-local signal due to effective pinning potentials and magnon scattering, which could compound the effects seen here. It should be noted that a hysteresis in the position of *n* can appear in systems without a bulk $H_D$, as recently observed in a CoO/Pt sample rotated in a magnetic field[33], but this hysteresis will disappear above the spin flop field whilst here it persists until $E_Z > E_D$. However, we clearly observe a hysteresis above $H_{sf}$, both in our experimental measurements of the magnon transport and the equilibrium position of *n* in the presence of DMI.

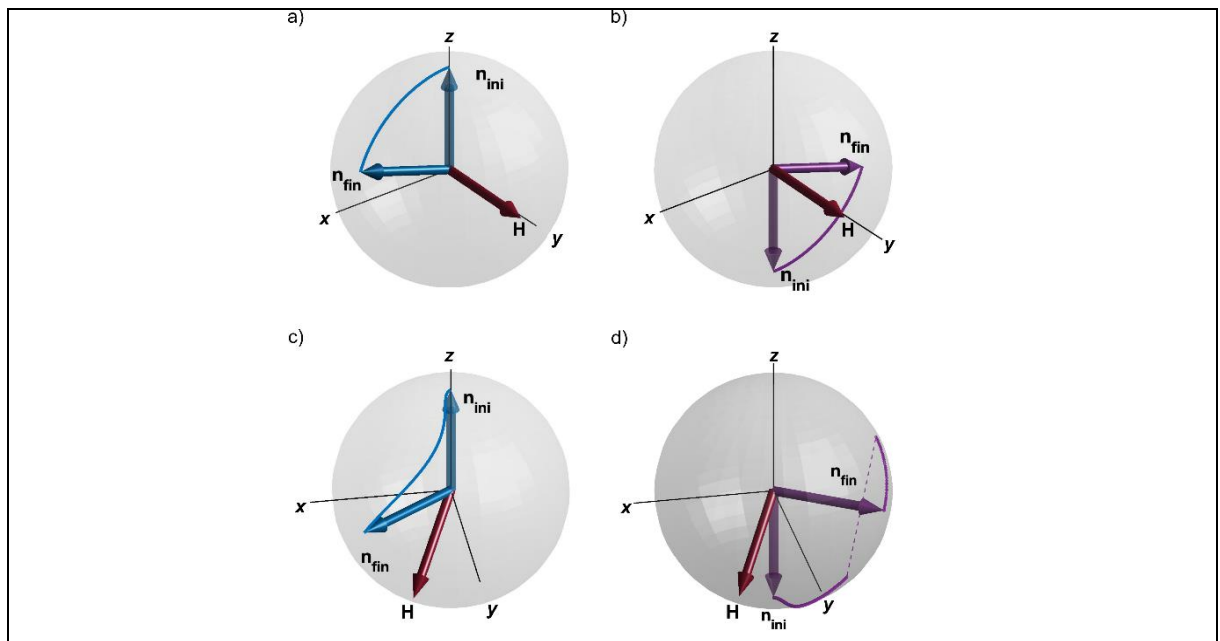

**Figure 4: Trajectory of the Néel vector under an applied field.** a) Trajectory of the Néel vector in hematite that initially lies along *z* when a magnetic field is applied along *y* when $H_D = 0$ T. b) Trajectory of the Néel vector in hematite that initially lies along -*z* when a magnetic field is applied along *y* when $H_D = 0$ T. The initial (translucent arrows) and final (solid arrows) of the Néel vector in a) and b) are mirror images. c) Trajectory of the Néel vector in hematite that initially lies along *z* when a magnetic field is applied along an arbitrary direction when $H_D \neq 0$ T. d) Trajectory of the Néel vector in hematite that initially lies along -*z* when a magnetic field is applied along an arbitrary direction when $H_D \neq 0$ T. The Néel vector in the presence of DMI follow a different trajectory depending on the initial configuration of *n* and *H*, where *n* may even show a discontinuity.

3. Conclusions

The effective field created by the Dzyaloshinskii-Moriya interaction in our thin films is determined to be 1.7 T ± 0.1 T, reduced from the value of $H_D$ found in bulk systems (2.1 T)[27]. We determine that the critical field for spin-flopping is reduced indicating that the effective anisotropy field directed along the crystallographic c-axis is correspondingly lower in our thin film system. Nevertheless, we have demonstrated that even this reduced Dzyaloshinskii-Moriya interaction in thin film hematite leads to a profound impact not only on the Néel vector direction but also on the magnon transport. A magnetic field rotated within the sample plane leads to a hysteresis due to a competition between the Dzyaloshinskii-Moriya interaction and the Zeeman energy under an applied magnetic field. This shows that the field of magnonic spin transport in antiferromagnets, with key contributions made by Sergio Rezende, is a complex area with exciting findings.

**Acknowledgements**

A.R. and M.K. acknowledge support from the Graduate School of Excellence Materials Science in Mainz (DFG/GSC 266). This work was supported by the Max Planck Graduate Center with the Johannes Gutenberg-Universität Mainz (MPGC). A. R., R. L. and M.K. acknowledge support from the DFG project number 423441604. R.L. acknowledges the European Union's Horizon 2020 research and innovation programme under the Marie Skłodowska-Curie grant agreement FAST number 752195. R.L. and M.K. acknowledge financial support from the Horizon 2020 Framework Programme of the European Commission under FET-Open grant agreement no. 863155 (s-Nebula) O.G. and J.S. acknowledge the Alexander von Humboldt Foundation, the ERC Synergy Grant SC2 (No. 610115). All authors from Mainz also acknowledge support from both MaHoJeRo (DAAD Spintronics network, project number 57334897 and 57524834), SPIN+X (DFG SFB TRR 173 No. 268565370, projects A01, A03, B02, and B12), and KAUST (OSR-2019-CRG8-4048.2). A. K. and Av.R. acknowledges support from the European Research Council under the European Union's Seventh Framework programme

(FP/200702013) / ERC (Grant Agreement No. 617516). D.A.G. acknowledges support from The Center for Absorption in Science, Ministry of Immigrant Absorption, State of Israel.